\documentclass[journal]{IEEEtran}
\usepackage{cite}
\usepackage{amsmath,amssymb,amsfonts}
\usepackage{algorithmic}
\usepackage{graphicx}
\usepackage{textcomp}
\usepackage{siunitx}
\usepackage{xcolor}
\usepackage{url}

\begin{document}
%
\title{Influence of Coil Position on AC Losses of Stator Superconducting Windings of a Synchronous Machine for a 10 MW Wind Turbine}

%

\author{Carlos~Roberto~Vargas-Llanos,
        Sebastian~Lengsfeld,
        Mathias~Noe,
        Tabea~Arndt, 
        and~Francesco~Grilli
\thanks{The underlying work of this article was funded by the German Federal Ministry for Economic Affairs and Energy (project name ``SupraGenSys'', funding reference number 03EE3010A and 03EE3010D). The responsibility for the content of this article lies with the authors and does not necessarily reflect the opinion of the SupraGenSys project consortium.}
\thanks{ C.R. Vargas-Llanos, M. Noe, T. Arndt and F. Grilli are with the Institute for Technical Physics (ITEP) of the Karlsruhe Institute of Technology (KIT), 76131 Karlsruhe, Germany (e-mail: francesco.grilli@kit.edu).}
\thanks{S. Lengsfeld is with Fraunhofer Institute for Energy Economics and Energy System Technology, 34119 Kassel, Germany}

}

\markboth{Journal of \LaTeX\ Class Files,~Vol.~X, No.~X, February~2021}%
{Shell \MakeLowercase{\textit{et al.}}: Bare Demo of IEEEtran.cls for IEEE Journals}

\maketitle

\begin{abstract}
One of the main tasks during the design of a superconducting electrical machine is the estimation of losses in the superconducting coils. These losses can be decisive in such applications since they influence the cooling power requirements and the overall efficiency of the machine. In this publication, we focus on the dissipation in the stator superconducting coils of a synchronous machine for a wind turbine application. The T-A formulation of Maxwell's equations is used in a 2D finite element model to analyse the behaviour of the magnetic field around the coils and calculate losses.
Particular attention is given to the position of the coils inside a slot and several coil configurations are presented. It is shown that certain coil arrangements lead to a significantly lower total loss, a more uniform loss distribution, which ultimately leads to the possibility of increasing the operating temperature. 

\end{abstract}


\IEEEpeerreviewmaketitle

\section{Introduction}\label{Intro}
\IEEEPARstart{T}{he} large current capacity of high-temperature superconductors (HTS) has inspired several applications. One of the main research topics is related to superconducting electrical machines for transport and energy conversion \cite{snitchler_10_2011}, \cite{moon_development_2016}, \cite{gamble_full_2011}, \cite{haran_high_2017}. In this field, superconductors can bring a significant reduction of the  size and the weight of the applications. Moreover, their zero resistance can increase the efficiency of the machine in comparison with conventional solutions \cite{kalsi_applications_2011}. This has encouraged several investigations and projects focusing on generators for wind turbine applications~\cite{abrahamsen_large_2012}, \cite{kalsi_superconducting_2014}, \cite{sanz_superconducting_2014}, \cite{liu_comparison_2015}, \cite{noauthor_ecoswing_nodate}, where the increasing power consumption and renewable energy generation goals have pushed towards more compact and efficient solutions. 

Despite their zero resistance, HTS experience losses under time changing current or magnetic field~\cite{wesche_physical_2015}. Moreover, the superconductor must be cooled at cryogenic temperatures to take advantage of its superconducting properties. Under such operating conditions, the cooling efficiency can increase the cooling power requirements. Therefore, the losses in the superconductor can be decisive for the design and construction of superconducting coils in electrical machines~\cite{zhang_ac_2013}. Most superconducting generators for wind power use the HTS wires/tapes to realize a more compact rotor, taking into consideration that the field winding carries DC-current. The airgap flux density is mainly unchanged, so existing copper stator designs can be kept. We propose in this investigation new arrangements of coils that allow us to integrate the superconductors in the stator and decrease losses in comparison with typical configurations.

Several analytical equations have been developed to estimate losses in  tapes and stacks of superconducting tapes \cite{mawatari_critical_1996}, \cite{mawatari_critical_1997}, \cite{mikitik_analytical_2013}. However, these expressions only represent specific operating conditions such as AC transport current or uniform external magnetic field. In this publication, we estimate losses in the stator's superconducting coils of a synchronous machine for a wind power application. Therefore, we follow the approach described in \cite{vargas-llanos_t-formulation_2020} to model the electrical machine by using the T-A formulation in a finite-element model. The basic design and a brief description of the modelling technique are presented in section  \ref{Machine_and_modelling}. A detailed analysis of the current penetration and the behaviour of the magnetic field around the coils is introduced in section \ref{Configurations}. This allows us to propose several configurations that can reduce the AC losses and achieve a better distribution of the average power density dissipation in the HTS coils. Section \ref{Summary_and_temperatures} summarizes the losses in all the designs and introduces a sensitivity analysis over different temperatures. Finally, the main findings of this work are summarized in section~\ref{Conclu}.






\section{Modelling Technique and Machine Characteristics}\label{Machine_and_modelling}

\subsection{Modelling Technique}

The T-A formulation of Maxwell's equations is used here to estimate losses in the superconducting tapes by following the approach described in~\cite{vargas-llanos_t-formulation_2020}. This means that the induced stator currents are estimated with a star connected load in the building model process. Then, this data is imported into the T-A formulation where the magnetic vector potential \textbf{A} is computed in the entire geometry as \cite{zhang_efficient_2016} \cite{liang_finite_2017}:

\begin{equation} \label{eq_Amp_A_formulation}
\nabla \times (\frac{1}{\mu} \nabla \times \mathbf{A}) = \mathbf{J}. \end{equation}   
where \textbf{J} is the current density, \(\mu\) is the permeability of the material and \textbf{A} is defined as \( \mathbf{B}=\nabla \times \mathbf{A}\) (\textbf{B}, magnetic flux density). The current vector potential \textbf{T} ( \(\mathbf{J}=\nabla \times \mathbf{T}\)) is only calculated in the superconducting domain as:

\begin{equation} \label{eq_T_formulation}
\nabla \times (\rho _{\rm HTS} \nabla \times \mathbf{T}) = -\frac{\partial \mathbf{B}}{\partial t}.
\end{equation}

The superconducting tapes are modelled with a non-linear resistivity \cite{rhyner_magnetic_1993}:
\begin{equation} \label{eq_HTS_resis}
\rho _{\rm HTS}=\frac{E_{\rm c}}{J_{\rm c}(\mathbf{B})}  \Bigg| \frac{\mathbf{J}}{J_{\rm c}(\mathbf{B})} \Bigg| ^{n-1}.
\end{equation}
 
In all the calculations, we considered  a critical electric field \(E_{\rm c}=\SI{1e-4}{\volt\per\meter}\) and a value of \(n=25\). Finally, the losses in the superconducting tapes \textbf{Q} are calculated by integrating the dot product between electric field \textbf{E} and current density \textbf{J} in the superconducting tape over half a cycle: 
 \begin{equation} \label{Losses_per_racet}
Q_{\rm HTS} (\SI{}{\joule\per\meter})= \frac{2}{T} \int_{T}^{\frac{3T}{2}} \int \mathbf{E} \cdot \mathbf{J} dl dt.
\end{equation}
 
The models used in this publication are a 2D representation of an electrical machine. Therefore, the end-effects are not considered. 

Since the electromagnetic behaviour is periodically repeated in space and time, one pole pair is analyzed and only one group of coils is modelled with details~\cite{vargas-llanos_t-formulation_2020}. Periodic boundary conditions allow us to rebuild the whole machine cross-section without affecting the accuracy of the calculation.

The total losses in the machine's stator coils are calculated by multiplying the estimation for one group of coils by the number of coil groups, the number of pole pairs and the active length of the machine.

\subsection{Description of the Electrical Machine}

\begin{figure}[hbt!]
\centerline{\includegraphics[width=0.4\textwidth]{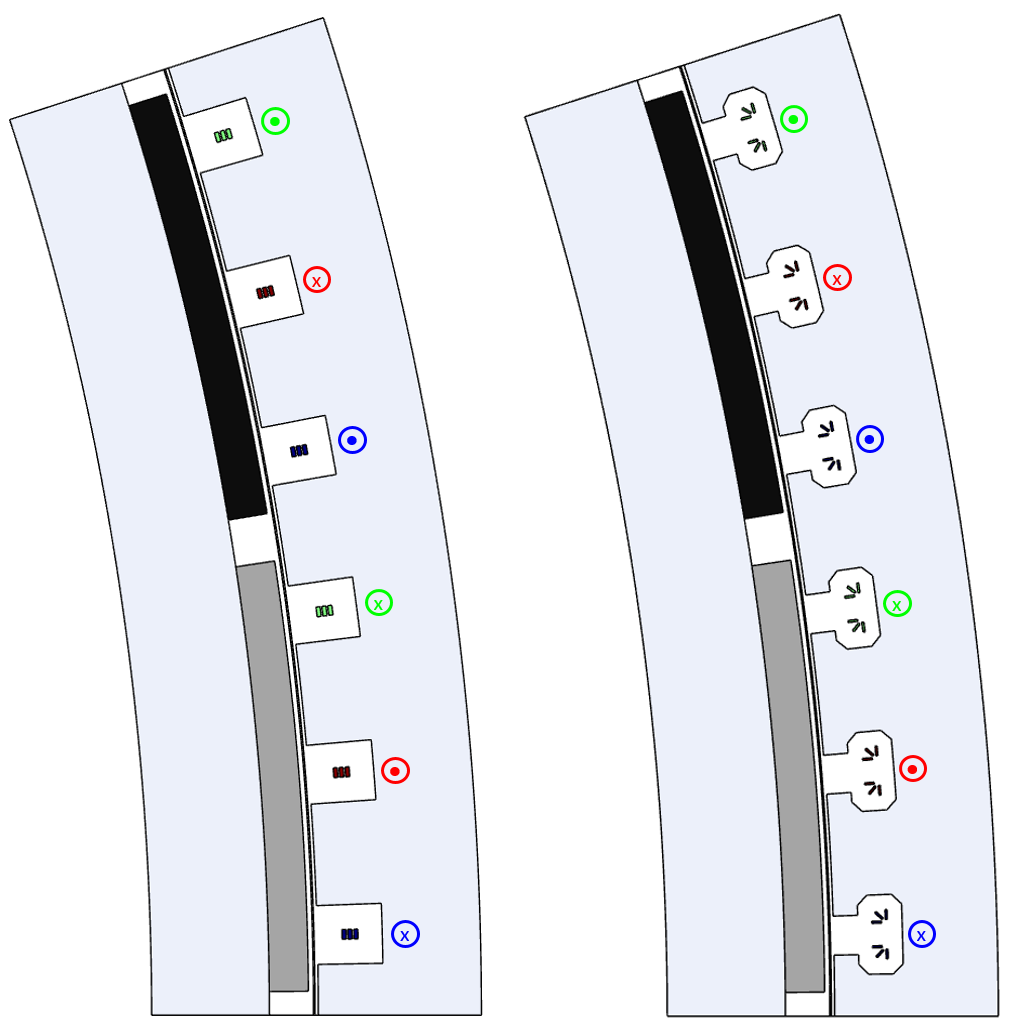} }
\caption{One pole pair cross-section of the machine design. From left to right: original and modified slot alternatives. Phase A is depicted in blue, B in red and C in green. The direction of the current is indicated in a circle close to the coils.}
\label{Geometry}
\end{figure}

The main design under study is a direct driven synchronous electrical machine with permanent magnets in the rotor and superconducting distributed winding in the stator. This machine has an active length of \SI{2.4}{\meter} to achieve a rated torque of \SI{10}{\mega\newton\meter}. The stator coils are located in rectangular slots in almost all the analyzed arrangements. This rectangular shape of the slot is modified in the final configuration to reduce losses and distribute the coils inside the slot. The cross-section of one pole pair is shown in Fig.~\ref{Geometry} and the main characteristics are summarized in table~\ref{table_gen}. 

The magnetic material is M330-50A, whose main properties are described in \cite{vargas-llanos_t-formulation_2020} and \cite{liu_measurement_2018}. This material is assumed to be working at room temperature since iron losses can increase under cryogenic temperatures.

The stator coils are wound with a \SI{4}{\milli\meter} REBCO tape with a critical current density of \SI{3}{\mega\ampere\per\square\centi\metre} at \SI{77}{\kelvin} and an average operating temperature of \SI{65}{\kelvin} \cite{hazelton_status_2013}, \cite{zou_simulation_2016}. The tape, and its non-linear dependence of the critical current density on the magnetic field amplitude and direction, is modelled as it was done in \cite{vargas-llanos_t-formulation_2020}.

\begin{table}[hbt!]
\centering
\caption{Main parameters of the superconducting generator.}
\label{table_gen}
\begin{tabular}{l c} 
\hline
\hline			
Number of slots	                        &	120	\\
Number of pole pairs                	&	20	\\
Number of slots per pole pair	        &	6	\\
Remanent induction of permanent magnets	&	\SI{1.28}{\tesla}	\\
Number of coils per phase per pole pair	&	3	\\
Number of turns per coil	            &	50	\\
Peak current in each coil	            &	\SI{141.4}{\ampere}	\\
Rated electrical frequency    &	\SI{3.33}{\hertz}	\\
Rated power                	&	\SI{10}{\MW}	\\
\hline
\hline
\end{tabular}
\end{table}


\section{Configuration Analysis and Estimation of Losses}\label{Configurations}
In this section, several configurations are studied by considering the same rectangular slot design. The coil's position and orientation inside the slot are changed based on the average power density dissipation and electromagnetic behaviour. This analysis is done at full load and  over the group of coils located in the first slot from bottom to top depicted in blue in Fig.~\ref{Geometry}. However, the same changes are implemented in all the coils and slots for each solution. 

Changes in the shape of the slot are introduced in the last analyzed configuration. These changes are done to locate the coils inside the slot, reduce losses and keep a similar arrangement between the \SI{4}{\milli\meter} and \SI{2}{\milli\meter} tape's width alternatives.

\subsection{Typical Arrangement}
\begin{figure}[hbt!]
\centerline{\includegraphics[width=0.3\textwidth]{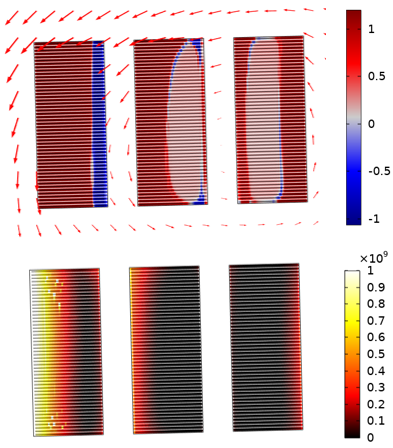} }
\caption{From top to bottom: behaviour of the normalized current density (\(J/J_{\rm c}(\mathbf{B})\)) and magnetic field (red arrows) when the current in the coil group is maximum and average power density dissipation (\SI{}{\W\per\cubic\meter}) in the typical arrangement with a \SI{4}{\milli\meter} tape.}
\label{TA_4mm}
\end{figure}

The first configuration to analyze uses coils with a racetrack form. This shape is considered typical in superconducting electrical machines. Three coil layers are stacked one on top of the other in rectangular slots for one phase in one pole pair. These coils are wound with a \SI{4}{\milli\meter} tape for an initial calculation. Fig.~\ref{TA_4mm} shows the behaviour of the normalized current density (\(J/J_{\rm c}(\mathbf{B})\)) and average power density dissipation for this alternative. We have calculated the average dissipation $P_{\rm avg}$ in each point of the superconducting tapes as: 

\begin{equation} \label{ave_power_density}
P_{\rm avg}=\frac{2}{T} \int_{T}^{\frac{3T}{2}} \mathbf{E} \cdot \mathbf{J} dt.
\end{equation}

Fig.~\ref{TA_4mm} shows a high current penetration in the first coil from left to right (closer to the rotor). This coil experiences a higher perpendicular component (to the flat face of the tape) of the magnetic field in comparison with the other two. We can also notice that most of the dissipation is located in the same coil. Losses are higher on the left side, where the magnetic field is higher. The total losses in the superconducting stator winding are \SI{28}{\kW}.    

\begin{figure}[hbt!]
\centerline{\includegraphics[width=0.3\textwidth]{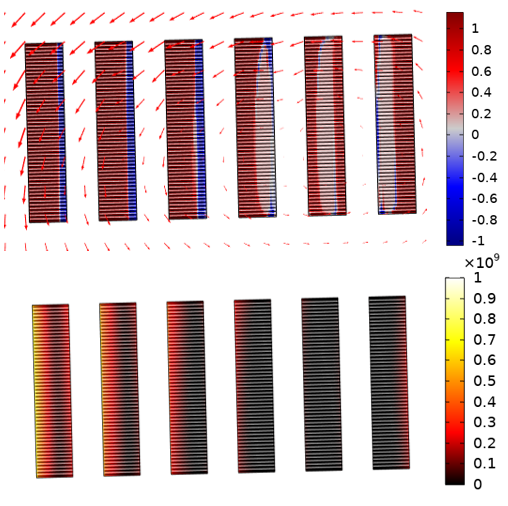} }
\caption{From top to bottom: behaviour of the normalized current density (\(J/J_{\rm c}(\mathbf{B})\)) and magnetic field (red arrows) when the current in the coil group is maximum and average power density dissipation (\SI{}{\W\per\cubic\meter}) in the typical arrangement with a \SI{2}{\milli\meter} tape.}
\label{TA_2mm}
\end{figure}

We can observe in Fig.~\ref{TA_2mm} the normalized current density (\(J/J_{\rm c}(\mathbf{B})\)) and average power density dissipation for the same arrangement with a \SI{2}{\milli\meter} tape. We have now six layers stack one on top of the other for each phase in one pole pair. In this calculation, we have used the same parameters of the \SI{4}{\milli\meter} HTS tape (only the transport and critical current are reduced to half). This figure shows a similar behaviour between \SI{2}{\milli\meter} and \SI{4}{\milli\meter} alternatives. We have higher current penetration and dissipation in the first coils from left to right (closer to the rotor). This uneven distribution of losses can be related to the high perpendicular component of the magnetic field in these coils. However, the total losses in the superconducting coils are \SI{15.42}{\kW}. 



\subsection{90-Degree Inclined Coils}
\begin{figure}[hbt!]
\centerline{\includegraphics[width=0.35\textwidth]{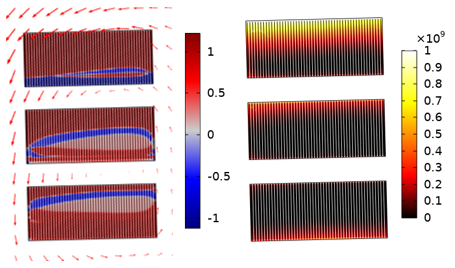} }
\caption{From left to right: behaviour of the normalized current density (\(J/J_{\rm c}(\mathbf{B})\)) and magnetic field (red arrows) when the current in the coil group is maximum and average power density dissipation (\SI{}{\W\per\cubic\meter}) in the 90 degrees inclined coils arrangement with a \SI{4}{\milli\meter} tape.}
\label{90_4mm}
\end{figure}

In this arrangement, we have rotated the cross-section of the coils 90 degrees in comparison with the previous solutions. Therefore, the tapes closer to the rotor are better aligned with the magnetic field. Fig.~\ref{90_4mm} shows the behaviour of the normalized current density (\(J/J_{\rm c}(\mathbf{B})\)) and average power density dissipation for the \SI{4}{\milli\meter} tape alternative. The first coil from top to bottom has now a high current penetration and it is exposed to a high perpendicular component of the magnetic field. Therefore, the highest average dissipation is located in this coil. The total losses for this alternative are \SI{22.85}{\kW}. 

\begin{figure}[hbt!]
\centerline{\includegraphics[width=0.35\textwidth]{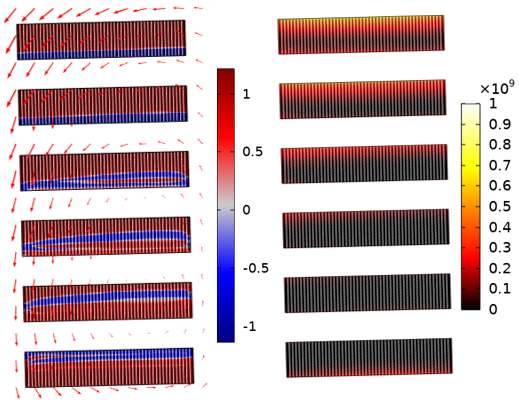} }
\caption{From left to right: behaviour of the normalized current density (\(J/J_{\rm c}(\mathbf{B})\)) and magnetic field (red arrows) when the current in the coil group is maximum and average power density dissipation (\SI{}{\W\per\cubic\meter}) in the 90 degrees inclined coils arrangement with a \SI{2}{\milli\meter} tape.}
\label{90_2mm}
\end{figure}

The behaviour of the normalized current density (\(J/J_{\rm c}(\mathbf{B})\)) and average power density dissipation for the same arrangement with a \SI{2}{\milli\meter} tape is shown in Fig.~\ref{90_2mm}. We have modelled the tape by following the same approach mentioned in the previous section.  The first coils from top to bottom are still experiencing the highest average power density dissipation due to a high perpendicular component of the magnetic field and a high current penetration. The total losses in the superconducting coils for this \SI{2}{\milli\meter} tape alternative are \SI{13.21}{\kW}.  

\subsection{Top and Bottom Inclined Coils}
\begin{figure}[hbt!]
\centerline{\includegraphics[width=0.4\textwidth]{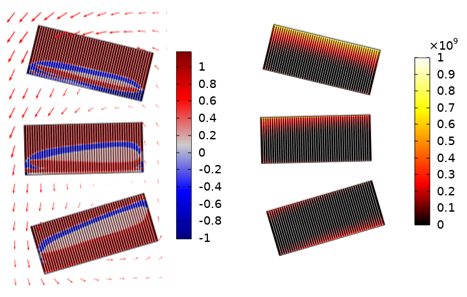} }
\caption{From left to right: behaviour of the normalized current density (\(J/J_{\rm c}(\mathbf{B})\)) and magnetic field (red arrows) when the current in the coil group is maximum and average power density dissipation (\SI{}{\W\per\cubic\meter}) in the top and bottom inclined coils arrangement with a \SI{4}{\milli\meter} tape.}
\label{TB_4mm}
\end{figure}

Since we still had an uneven distribution of losses in the last arrangement, we have decided to incline the top and bottom coils. This approach will allow us to reduce the perpendicular component of the magnetic field around the coils, which will further reduce the losses. In Fig.~\ref{TB_4mm} we can appreciate the behaviour of the normalized current density (\(J/J_{\rm c}(\mathbf{B})\)) and average power density dissipation for the \SI{4}{\milli\meter} tape alternative. As it can be seen, the current penetration and average power density dissipation have decreased in comparison with the previous arrangement. The total losses in the superconducting coils for this configuration are \SI{14.73}{\kW}.  

\begin{figure}[hbt!]
\centerline{\includegraphics[width=0.4\textwidth]{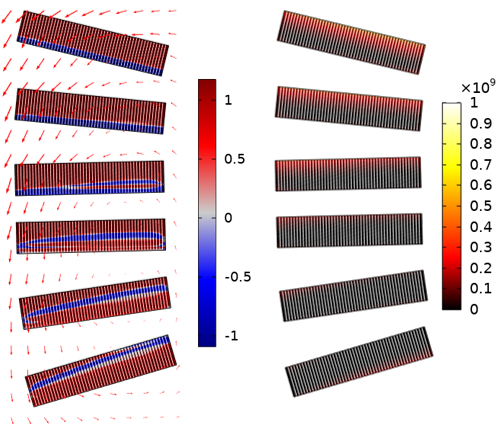} }
\caption{From left to right: behaviour of the normalized current density (\(J/J_{\rm c}(\mathbf{B})\)) and magnetic field (red arrows) when the current in the coil group is maximum and average power density dissipation (\SI{}{\W\per\cubic\meter}) in the top and bottom inclined coils arrangement with a \SI{2}{\milli\meter} tape.}
\label{TB_2mm}
\end{figure}

Fig.~\ref{TB_2mm} shows the behaviour of the normalized current density (\(J/J_{\rm c}(\mathbf{B})\)) and average power density dissipation for the \SI{2}{\milli\meter} tape alternative. We can notice from a direct comparison between Fig.~\ref{TB_2mm} and Fig.~\ref{90_2mm} that the average power dissipation in the first coils from top to bottom has decreased. This behaviour can be related to the alignment of the tapes with the magnetic field. The total losses in the superconducting coils for this \SI{2}{\milli\meter} tape design are \SI{10.3}{\kW}.  

\subsection{Star Configuration}
\begin{figure}[hbt!]
\centerline{\includegraphics[width=0.3\textwidth]{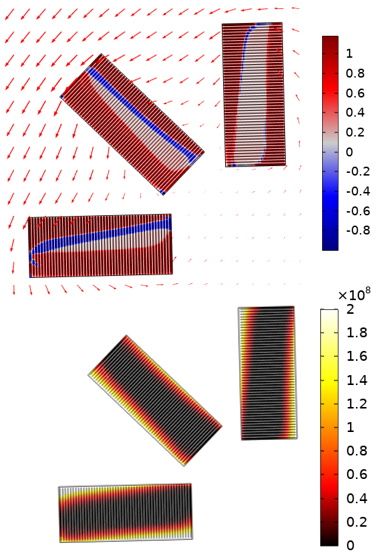} }
\caption{From top to bottom: behaviour of the normalized current density (\(J/J_{\rm c}(\mathbf{B})\)) and magnetic field (red arrows) when the current in the coil group is maximum and average power density dissipation (\SI{}{\W\per\cubic\meter}) in the star arrangement with a \SI{4}{\milli\meter} tape.}
\label{Star_4mm}
\end{figure}

We have noticed in the previous configuration analysis that the orientation of the tape can play a key role in the average power density dissipation. For this reason, we propose a new arrangement of the cross-section of the coils by trying to align more the tapes with the magnetic field. This approach allows us to reduce further the perpendicular component of the magnetic field. Fig.~\ref{Star_4mm} shows the behaviour of the normalized current density (\(J/J_{\rm c}(\mathbf{B})\)) and average power density dissipation for the \SI{4}{\milli\meter} tape alternative of this new star configuration. We can observe in this figure that the current penetration is now similar between coils, and the dissipation is better distributed. Furthermore, the total losses in the superconducting winding are \SI{5.57}{\kW}. 

\begin{figure}[hbt!]
\centerline{\includegraphics[width=0.48\textwidth]{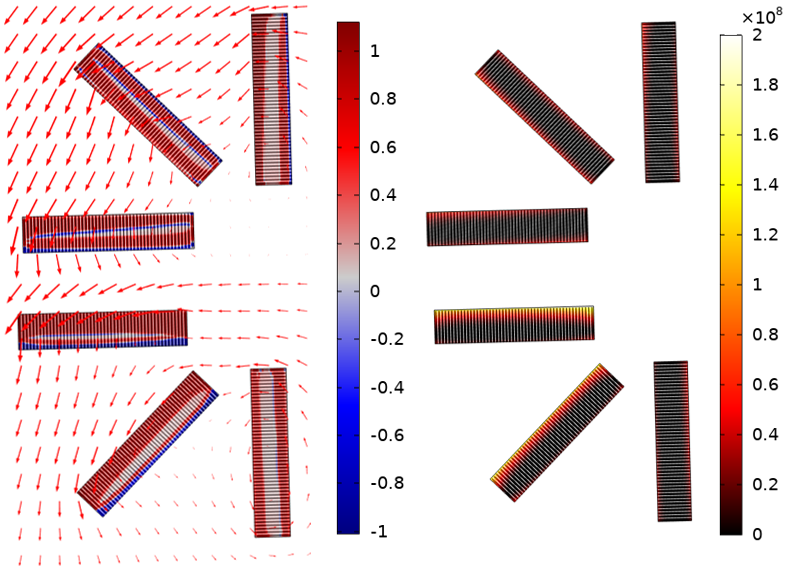} }
\caption{From left to right: behaviour of the normalized current density (\(J/J_{\rm c}(\mathbf{B})\)) and magnetic field (red arrows) when the current in the coil group is maximum and average power density dissipation (\SI{}{\W\per\cubic\meter}) in the star arrangement with a \SI{2}{\milli\meter} tape.}
\label{Star_2mm}
\end{figure}

A reduction in losses has been achieved in the previous configurations by decreasing the tape's width. Therefore, we have followed the same approach for this design. The shape of the slot has been modified to locate the coils in the slot,  keep the same arrangement and further improve the electromagnetic behaviour \cite{vargas-llanos_t-formulation_2020}. We can observed in Fig.~\ref{Star_2mm} the behaviour of the normalized current density (\(J/J_{\rm c}(\mathbf{B})\)) and average power density dissipation for this \SI{2}{\milli\meter} tape alternative. The total losses in the superconducting winding for this configuration are \SI{4.15}{\kW}.

\section{Summary of Losses and Behaviour for Different Temperatures}\label{Summary_and_temperatures}

The total losses in the superconducting coils are summarized in table~\ref{table_total_losses}. We can notice in this table that a continuous reduction of losses is achieved through the different arrangements. This loss reduction can be classified under two main strategies.

The first one is a reduction in the tape's width. This approach can bring a reduction in the magnetization losses of single tapes under a uniform perpendicular magnetic field \cite{brandt_type-ii-superconductor_1993} and brings a \SI{45}{\percent}  reduction in losses for the typical and a \SI{30}{\percent} reduction in losses for the top and bottom arrangement. However, the average power density dissipation keeps an uneven distribution. This can cause higher thermal stress in some coils during normal operation.  

The second strategy is based on the magnetic field orientation. Therefore, the position of the coils was changed to decrease the perpendicular component of the magnetic field. If we consider the typical arrangement as a reference, this approach brings a \SI{80}{\percent} reduction in losses for the star configuration with a \SI{4}{\milli\meter} tape, and a \SI{73}{\percent} reduction in losses for the star configuration with a \SI{2}{\milli\meter} tape. Moreover, the distribution of the average power density dissipation is better. This represents a better distribution of the thermal stress  during normal operation. 

\begin{table}[hbt!]
\centering
\caption{Total losses in the HTS coils at \SI{65}{\K}, electric stator frequency$=\SI{3.33}{\hertz}$ and rated load.}
\label{table_total_losses}
\begin{tabular}{c c c c c} 
\hline
\hline	
Tape width & \multicolumn{4}{c}{Configuration} \\
\hline
 & Typical & 90 degrees & Top and bottom & Star \\
\SI{4}{\milli\meter} & \SI{28}{\kW} & \SI{22.85}{\kW} & \SI{14.73}{\kW} & \SI{5.57}{\kW} \\
\SI{2}{\milli\meter} & \SI{15.42}{\kW} & \SI{13.21}{\kW} & \SI{10.3}{\kW} & \SI{4.15}{\kW} \\
\hline
\hline
\end{tabular}
\end{table}

\begin{figure}[hbt!]
\centerline{\includegraphics[width=0.5\textwidth]{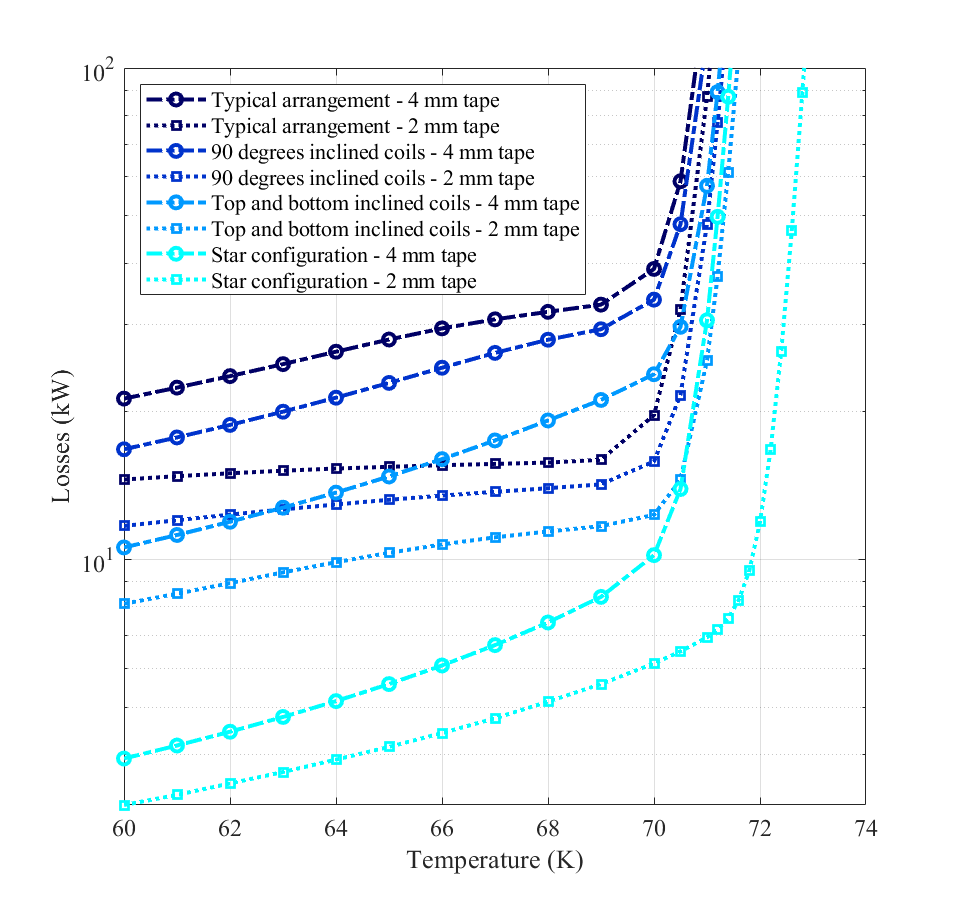} }
\caption{Behaviour of the losses in the HTS coils for different temperatures and rated frequency of \SI{3.33}{\hertz}.}
\label{losses_vs_temp}
\end{figure}

Fig.~\ref{losses_vs_temp} shows the results of a parametric study to investigate the behaviour of total losses in the HTS coils for different average temperatures. We  notice in this plot that decreasing the tape's width can allow us to increase the operating temperature. A further increase in temperature is feasible in the star configuration. This advantage can be related not only to the decrease in the losses but also to the better distribution of the dissipation. Therefore, this arrangement offers a more secure and efficient operation of the superconducting coils. 

\section{Conclusion}\label{Conclu}

Losses have been estimated in the HTS coils of a synchronous machine with permanent magnets in the rotor and superconducting winding in the stator. We have used the T-A formulation of Maxwell's equations to model the generator by employing finite elements. 
A detailed analysis of the electromagnetic behaviour of the coils at rated conditions has been done, several configurations have been proposed and a continuous reduction in losses has been achieved. The loss reduction has been  classified into two main strategies based on the tape's width and alignment. This last approach has provided up to \SI{80}{\percent} reduction in losses and a better average power density dissipation in the HTS coils of the synchronous machine under study. The analysis has been  complemented with a sensitivity study over different temperatures. 
From all the analyzed arrangements, the star configuration offers the lowest losses and the best average power density  distribution. Moreover, it enables an increase in the operating temperature and a better distribution of the thermal stress due to a better distribution of the losses. 






\section*{Acknowledgment}
 The authors thank Dr. Marijn Pieter Oomen (Siemens AG) for sharing comments regarding superconducting electrical machines that fed fruitful discussions during the development of this work.

\bibliographystyle{IEEEtran}
\bibliography{My_Library_03022021.bib}

\end{document}